\documentstyle[12pt]{article}
\begin{document}
\begin{titlepage}
\begin{center}
EUROPEAN ORGANIZATION FOR NUCLEAR RESEARCH
\end{center}
\vglue1cm
\begin{flushright}
CERN-LAA/94-26\\
CERN-TH.7373/94
\end{flushright}
\vglue1cm
\begin{center}{\bf EXPLICIT SUPERSTRING VACUA IN A BACKGROUND
 OF GRAVITATIONAL WAVES AND DILATON}
\vglue 1cm
A. Peterman$^{\rm a,b)}$ and A. Zichichi$^{\rm b)}$
\vglue.5cm
$^{\rm a)}$ CNRS-Luminy, Marseilles, France\\
$^{\rm b)}$ CERN, Geneva, Switzerland
\vglue3cm

{\bf Abstract}\\
\end{center}

 We present an explicit solution of superstring effective equations,
represented by gravitational waves and dilaton backgrounds.  Particular
solutions will be examined in a forthcoming note.

\vglue4cm
\begin{flushleft}
CERN-LAA/94-26\\
 CERN-TH.7373/94\\
 July 1994
\end{flushleft}
\end{titlepage}

\section{Introduction}

 In the conclusions of
a recent paper on supersymmetric non-linear sigma models with Brinkmann
metric
\cite{peterm1}, we announced the possibility to display explicit metric
and dilaton
backgrounds which satisfy the Weyl conditions.  The assumption, for this
result
to obtain, was that the transverse space of the model be hyperK\"ahler,
therefore
implying for this model an N = 4 supersymmetry.  Such supersymmetric
model has
been worked out explicitly. It provides the backgrounds for both the
metric and
the dilaton in closed form and is the purpose of this short
communication.

 After
a brief outline of the hyperK\"ahler manifold chosen for the transverse
space of
the model, we give the metric in this space, that was taken in the
Calabi series
\cite{peterm2}, i.e. a cotangent bundle over CP$^1$: T(CP$^1$)$^*$. The
background for this model
is notoriously self-dual: of the Eguchi-Hanson type.  At this point the
crucial
feature is that N = 4 supersymmetric sigma-models (requiring a manifold
which is
hyperK\"ahler), are known \cite{peterm3} to have an identically zero
beta-function\footnote{Notice that preservation of N = 4 sypersymmetry
in perturbation
theory requires that the metric plus eventual counterterms preserve the
Ricci-flatness
\cite{peterm4}.}.
  Accordingly the
Weyl invariant background for the metric is equal, up to a
multiplicative
constant\footnote{In principle the sigma-model inverse
constant $f$ might be a
complicated function of the (n + 1)$^{\rm  th}$ coordinate u.
Here it is constant.}, to
the original metric one started with.
Moreover, the generic reparametrization term M
reduces to M$\mu =
\partial_\mu\phi$ since W$\mu$ can be shown to be zero
for such models having vanishing
beta-function of their transverse part.

Finally, as these hyperK\"ahler spaces are necessarily Ricci-flat
\cite{peterm4},
the dilaton may
be reduced to its linear form in the 0$^{\rm th}$ and (n + 1)$^{\rm th}$
coordinates of the model,
i.e. the light cone coordinates u and v.

 As conclusions and perspectives, we call the reader's
attention on the following fact: metric on various hyperK\"ahler
manifolds are seldom explicitly
known, although, for most of them, existence theorems can be proven (for
instance the celebrated
K3-spaces).  Recent studies, however, have considerably enlarged the
field \cite{peterm5}.  The
progresses in this direction should allow to formulate whole series of
valuable models with non
trivial space-time. The existence of an N = 4 world sheet superconformal
symmetry seems to have a
stabilizing effect on the perturbative solution.  Actually we refer to
the advances made in
\cite{peterm5} on superstrings in wormhole-like backgrounds.

The basic features of the
supersymmetric non-linear sigma-model with Brinkmann metric, Minkowski
signature and covariantly
constant null Killing vector, will of course not be recalled here, as a
series of rather
exhausting papers on the subject have appeared these two last years
\cite{peterm6}--\cite{peterm9}.  We recall the general form of the line
element for a
suitable choice of coordinates
\begin{equation}
\rm 	ds^2 = g_{\mu\nu} dx^\mu dx^\nu = -2 du \,dv + g_{ij} (x,u) dx^i
dx^j
\end{equation}
$\mu,\,\nu$ = 0,  1,..., n,
n + 1;  i,  j = 1,  2,...,  n (real indices).  The transverse part (with
latin indices) has a
metric that we will choose in such a way as it complies with the
manifold
described above on which transverse strengths take values.  We will
start, to
alleviate the computation by assuming that $\rm g_{ij}(x,u)$ can be
brought to the form
\begin{equation}
\rm g_{ij}(x,u) = {\it f}(u) g_{ij}(x)
\end{equation}
as was done in \cite{peterm6}--\cite{peterm9}.  This by no means
prohibits a
possible but more involved treatment of the problem by invoking further
additional symmetries which we shall not need to take into account here.
The
function $f$(u) in (2) can be shown to be the inverse coupling of the
supersymmetric transverse sigma-model and is therefore defined for a
family of
theories with various values of u.  It can be shown that $f$(u) is
running with u
and will satisfy a RG-like equation
\begin{equation}
	\rm p\dot{\it f}(u) = \beta {\it f}
\end{equation}
with p a constant and $\beta(f)$ the beta-function of the
transverse sigma-model, defined by the transverse $\beta{\rm G}_{\rm
ij}$
\begin{equation}
\beta^{\rm G}_{\rm ij} = \beta(f) \gamma_{\rm ij}\;.
\end{equation}
Finiteness of the model on a flat 2d background
requires in addition that the n + 2 sigma-model with target space metric
$\rm g_{\mu\nu}$
beta-function has to vanish up to a reparametrization term
D$_\mu$M$_\nu$.  M$_\mu$ is not
arbitrary, and to establish Weyl invariance of the model, the existence
of an
adequate dilaton background $\phi$ must be proved, such that M$_\mu$ is
represented by
\begin{equation}
	\rm M_\mu = \alpha^\prime\partial_\mu\phi + 1/2\;W_\mu\;.
\end{equation}

\section{The explicit model}

 The origin of W$_\mu$
has been discussed in several papers (see \cite{peterm6}--\cite{peterm9}
and \cite{peterm11} for
instance) and we do not repeat here the information we have on it.
Similarly we do not repeat
either the N = 1 supersymmetric extension \cite{peterm8,peterm9} of the
model with
bosonic action
\cite{peterm8,peterm9}.

In previous works on supersymmetric models of the kind studied in this
note, the number of supersymmetries considered for the transverse part
was N = 2.
Accordingly, K\"ahler manifolds were considered and especially symmetric
and
homogeneous K\"ahler manifolds.  These had the properties that the
beta-function
in (3) reduced to a constant and, by integration, $f$(u) was a constant
times u,
i.e.
$$f {\rm (u) = bu}$$
 (K\"ahler transverse space, homogeneous and symmetric).

In the
present note on the same class of models, we assume the transverse space
to be
an hyperK\"ahler manifold.  We will suppose the reader somewhat familiar
with
differential geometry and Lie groups in order to avoid to discuss here
the
holonomy group of such manifolds, in particular the absence of a U(1)
factor in
the holonomy group\footnote{The generator of this U(1) factor is the
Ricci-form
$\rm R^\alpha_{\alpha\gamma\bar\delta} dz\gamma\wedge d\bar
z^{\bar\delta}$ (complex indices).}
being the signature of a Ricci-flat manifold.  Also we suppose the
reader acquainted with the
Clifford algebra fulfilled by their three complex structures.  Anyhow,
one of the main features
that must be kept in mind is that hyperK\"ahler manifolds are
K\"ahlerian and quaternionic.  For
concreteness we specify the metric of our transverse space, as explained
at the beginning of this
note, to be that of a cotangent bundle over CP$^1$: T*(CP$^1$).  The
local hyperK\"ahler structure
of this space is proved by observing three linearly independent
covariantly constant complex
structures (or equivalently three closed 2-forms on T*(M) and show their
interrelation through
the SU(2) $\approx$  Sp(1) group transformations.  This can be found in
the literature
\cite{peterm12}--\cite{peterm14}.

Another feature to take into account is that, as can be seen
by construction, a higher symmetry of the metric on T*(M) is guaranteed
or at least coinciding
with the symmetry G of the initial manifold M, here CP$^1$.  The latter
is known to be a
symmetric space isomorphic to SU(2)/U(1) (or SO(3)/SO(2)).  With our
example
(T*(CP$^1$)) we automatically specialize to a transverse space which is
symmetric.
Then we can appreciably simplify the analysis\footnote{Rigorously
speaking, we have
already implicitly used this symmetry property in writing (3) and (4),
although parent
relations can be written down for non symmetric spaces at the price of a
much more
involved analysis (for comments see, for example, Ref. \cite{peterm8}).
Personally we
did not try to examine this case at this moment.}.  In particular, there
are both:
 1) restrictions on the form of the dilaton
potential which must be independent of the transverse space coordinates
and can be written as
\begin{equation}
\rm \phi = pv + \phi (u)
\end{equation}
 and 2) non-renormalization theorems for the dilaton \cite{peterm10}
which were already true at the level of N = 2 supersymmetry (i.e.
K\"ahler
transverse spaces), studied in \cite{peterm8} and \cite{peterm9} and
which can be shown to be a
{\it fortiori}  valid for hyperK\"ahler spaces like the one appearing in
the present
note.  This property considerably simplifies the Weyl invariance
differential
equation which was, in the N = 2 case
\begin{equation}
\rm \dot\phi ={A\over p} {\it f}^{-1}\beta ({\it f}) - {W_u\over 4} +
q\;;
\end{equation}
 A, p, q constants, $f$
constant, as $\beta (f)$ = 0 (q = 0 for critical D = 10).

We have explicitly shown in \cite{peterm9} that W$_{\rm u}$ was not
vanishing on homogeneous
K\"ahler spaces, due to the role of $f$(u) = bu in raising and lowering
the indices by
$\rm g_{ij}(x,u) = {\it f}(u) \gamma_{ij}(x)$ and its inverse, so that
W$_{\rm u}$ is proportional
to $\rm \partial_uS(u) \neq 0$, S(u) being the well-known globally
defined trace of curvature
power series (see also \cite{peterm15}).  In the present case $f$ is a
constant and as such cannot
introduce u dependence.  W$_{\rm u}$, like the other components of W  is
vanishing in the
present instance.  Furthermore the first term in (7) right-hand side is
vanishing as $\beta (f)$ does, and finally the answer for the dilaton
backgrounds reads
\begin{equation}
\rm\phi = \phi (u,v) = pv + qu + \phi_0
\end{equation}
($\phi_0$ constant) as announced at the very beginning
of this note.

What remains now is to display the line element in this model.  As
mentioned we took it as the simplest case (n = 1) in the Calabi series.
That
means we have to display the metric for a cotangent bundle over S$^2$ =
CP$^1$.  This
metric exists in the literature at several places (e.g. \cite{peterm2}).
Taking complex
coordinates for convenience and short-hand notations, the metric on any
T*(M)
has a block-form
$$\rm g_{T^* M} = \left(\matrix{\rm A&\rm B\cr \rm  C&\rm D}\right)$$
 with complex n$\times$n
matrices A, B, C, D, expressed in particular through the K\"ahler metric
on M:
\begin{equation}
\rm g_{\alpha\bar\beta} = \partial_\alpha\partial_{\bar\beta} K (z1,\bar
z1)\;;
\end{equation}
 $\rm (z1^\alpha  (\alpha = 1,...,$ n), a n component
complex variable) $\rm K(z1,\bar z1)$ being the K\"ahler potential.

One denotes another
2n real coordinates in cotangent space through $\rm z2_\alpha (\bar
z2_{\bar\alpha},\;(\alpha = 1,...,$ n).  In our case, with M = CP$^1$, n
= 1, we have
a 4-dimensional cotangent bundle T*(CP$^1$) as the transverse space in
(1).  Out of
these variables, the following scalar quantity can be defined
\begin{equation}
\rm t= g^{\alpha\bar\beta} (z1,\bar z1) z2_\alpha \bar z2_{\bar\beta}
\end{equation}
 and the application
\begin{equation}
\rm \Delta (t) = [-1+(1+4t)^{1/2}]/2t
\end{equation}
is a solution of the Ricci-flatness condition
$$\rm det(g_{T^*M}) = det \left(\matrix{\rm A&\rm B\cr\rm C&\rm
D}\right) = constant$$
with adequate
normalizations and regular behaviour at infinity.  One must say that it
is
remarkable that the determinant does not generally depend on z1 and z2
separately, but well on t only.  This feature allows one to deduce the
solution
(11).  It is necessary for (11) to obtain.  So, a candidate for a
K\"ahler
potential on T*(CP$^1$) can be guessed in the form
\begin{equation}
\rm K_{T^*M} (z1,\bar z1, z2,\bar z2) = K(z1,\bar z1) +{\cal{I}} (t)
\end{equation}
 In doing the
calculations of det(g$_{\rm T^*M}$), it appears soon that the candidate
(12) is improper
for most of the K\"ahler manifolds M in T*(M).  But it appears also that
in the
Calabi series (M = CP$^{\rm n}$), (12) is the right form to consider and
in particular
for T*(CP$^1$),
\begin{equation}
\rm \Delta (t) = {\cal {I}}^\prime (t) \equiv {d{\cal{I}} (t)\over
dt}\;.
\end{equation}
 After some
algebra, we can write the line element for the transverse space in the
following form
(depending on g$_{\alpha\bar\beta}$, its inverse, $\cal{I}$(t), and
derivatives with
respect to t, and on t itself of course:
\begin{equation}
\rm ds^2_T = A_{\alpha\bar\beta} dz1^\alpha d\bar z1^{\bar\beta} +
B^{\bar\beta}_\alpha
dz1^\alpha d\bar z2_{\bar\beta} + C^\alpha_{\bar\beta} d\bar z
1^{\bar\beta}
dz2_\alpha + D^{\alpha\bar\beta} dz2_\alpha d\bar z 2_{\bar\beta}
\end{equation}
with A, B, C and D given in appendix. Therefore (1) becomes
\begin{equation}
\rm ds^2 = -2 \,du\,dv+ {\it f} ds^2_T;\quad\quad {\it f}\;constant\;.
\end{equation}
 Hence, the resulting backgrounds
(14), ((15) below) and (8) represent exact explicit solutions of
superstring
effective equations, the so-called fixed point direct product solutions.
What is given
below can be found at several places in the literature.  It seemed to
the author
convenient for the reader to have this material close at hand.

\section{Mathematical Appendix}
 1) A, B, C and D appearing in formula (13) of the text
read
\begin{eqnarray}
\rm
A_{\alpha\bar\beta}&=&\rm  g_{\alpha\bar\beta} + \Delta\cdot
R_{\alpha\bar\beta\gamma\bar\delta} z 2^\gamma \bar z 2^{\bar\delta} +
\Gamma^\mu_{\alpha\gamma} z2_\mu D^{\gamma\bar\delta} \Gamma^{\bar
\nu}_{\bar\beta\bar\delta} \bar z2_{\bar\nu}\cr
&&\cr
\rm D^{\alpha\bar\delta} &=& \rm \Delta\cdot g^{\alpha\bar\beta} +
\Delta^\prime \cdot
z2^\alpha \bar z 2^{\bar \beta}\cr
&&\cr
\rm B^{\bar\beta}_\alpha&=& \rm \Gamma^\mu_{\alpha\gamma} z 2_\mu
D^{\gamma\bar\beta}\cr
&&\cr
\rm C^\alpha_{\bar\beta} &=&\rm  D^{\alpha\bar\delta}\cdot
\Gamma^{\bar\gamma}_{\bar\beta\bar\delta} \bar z 2_{\bar\gamma}\cr
&&\cr
\Delta&\equiv&\rm \Delta (t);\;\Delta^\prime \equiv{d\Delta(t)\over dt}
\end{eqnarray}
$\rm g_{\alpha\bar\beta}
,\;g^{\alpha\bar\beta},\;\Gamma^\alpha_{\beta\gamma},\;
R^\alpha_{\beta\gamma\bar\delta}$ are respectively the hermitian metric,
its inverse,
the connection and the curvature tensor on CP$^{\rm n}$.  $\Delta$ is
given by (11) in
the text.

2) Complex conjugation:
$$
\rm \overline{A_{\alpha\bar\beta}} = A_{\bar\alpha\beta};\quad\quad
\overline{D^{\alpha\bar\beta}} =
D^{\bar\alpha\beta};\quad\quad\overline{B^{\bar\beta}_\alpha} =
C^\beta_{\bar\alpha}
$$

3) Conditions for the metric on T*(CP$^{\rm n}$) to be hyperK\"ahler:
\begin{eqnarray}
\rm D^{\alpha\bar\beta} A_{\bar\beta\gamma} - C^\alpha_{\bar\beta}
B^{\bar\beta}_\gamma
&=& \rm \delta^\alpha_\gamma\cr
&&\cr
\rm C^\alpha_{\bar\beta} D^{\bar\beta\gamma} &=&\rm  D^{\alpha\bar\beta}
C^\gamma_{\bar\beta}\cr
&&\cr
\rm B^{\bar\beta}_\alpha A_{\bar\beta\gamma} &=&\rm  A_{\alpha\bar\beta}
B^{\bar\beta}_\gamma
\end{eqnarray}
	 plus three analogous conditions, got from (16) by complex conjugation.

4) CP$^1$
case.

The indices $\alpha ,\;\beta,...,\;\bar\alpha,\;\bar\beta,...$ can be
dropped, as we
have only two complex variables z1, z2 and their complex conjugates $\rm
\bar
z1, \bar z2$.  Using the explicit values of the metric, the connection
and
curvature tensor for the M = CP$^1$ case, the transverse line element
can be cast in the
simple form
\begin{equation}
\rm ds^2_T = G_{z1\bar z1} |dz1|^2 + G_{z1\bar z2}d\bar z1 dz2+ G_{\bar
z1z2} dz1d\bar
z2 + g_{\bar z2z2} |dz2|^2
\end{equation}
with the G's given by
\begin{eqnarray}
\rm G_{z1\bar z1} &=& \rm {1+4t(1+|z1|^2)\over \sqrt{1+4t}\cdot
(1+|z1|^2)^2}\cr
&&\cr
\rm G_{z1\bar z2}&=&\rm  \overline{G}_{\bar z1z2} = {2\bar
z1z2(1+|z1|^2)\over
\sqrt{1+4t}}
\cr
&&\cr
\rm G_{\bar z2z2}&=&\rm {(1+|z1|^2)^2\over \sqrt{1+4t}}
\end{eqnarray}

\end{document}